\documentstyle[aps,prl,twocolumn,psfig,floats]{revtex}
\newcommand{\be}{\begin{eqnarray}}
\newcommand{\ee}{\end{eqnarray}}
\newcommand\del{\partial}
\begin{document}
\draft
\wideabs{
\title{Random matrix triality at nonzero chemical potential}
\author{M.A. Halasz,  J.C. Osborn, and J.J.M. Verbaarschot}
\address{%
 Department of Physics, SUNY, Stony Brook, New York 11794}
\date{\today}
\maketitle
\begin{abstract}
We introduce three universality classes of
chiral random matrix ensembles with a nonzero chemical potential
and real, complex or quaternion real matrix elements.
In the thermodynamic limit we find that the distribution of the
eigenvalues in the complex plane does not depend on the Dyson index, and
is given by the solution proposed by Stephanov.
For a finite number  of degrees of freedom, $N$, we find 
an accumulation of eigenvalues on the imaginary axis for real 
matrices, whereas for
quaternion real matrices we find a depletion of eigenvalues in this domain.
This effect is of order $1/\sqrt N$. In particular for the real
case the resolvent shows a discontinuity of order $1/\sqrt N$. These
results are in agreement with lattice QCD simulations with
staggered fermions  and recent instanton liquid simulations 
both for two colors and a nonzero chemical potential.
\end{abstract}
}

\narrowtext
Recently, nonhermitean random matrices with eigenvalues scattered
in the complex plane have received a great deal of
attention in both condensed matter \cite{Sommer,Khor,fyodorov,Hatano,Efetov}
and QCD \cite{Stephanov,confirm}. In condensed matter they have been used
in problems ranging from neural networks \cite{Sommer} to the depinning 
transition of vortices in disordered superconductors \cite{Hatano,Efetov}.
In QCD,  they are relevant
at nonzero chemical potential when the Dirac operator is nonhermitean. 
In this letter we introduce three universality classes
of nonhermitean random matrices with the global symmetries of the QCD partition 
function. This allows us to interpret
certain characteristic features of quenched Dirac spectra at nonzero chemical
potential (i.e. with the fermion determinant ignored 
in generating the field configurations)
as signatures of anti-unitary symmetries. 
Such signatures have been found in condensed matter applications as well
\cite{Khor,fyodorov,Efetov}.

In quenched simulations of QCD it appears that the critical 
chemical potential
$\mu_c$ is proportional to the pion mass which vanishes in the
chiral limit as $\sqrt m$ ($m$ is the quark mass)
rather than to the baryon 
mass which remains nonzero for $m\rightarrow 0$  \cite{everybody,mendel}.
This long-standing puzzle has been resolved in a first successful application
of random matrix theory to QCD at nonzero chemical potential 
\cite{Stephanov}. The explanation \cite{Gocksch,Stephanov} is that the quenched 
limit is the limit $N_f \rightarrow 0$ of a partition function with the
fermion determinant to the power $N_f$
replaced by its absolute value. Such partition function
develops a condensate with Goldstone bosons consisting of a quark and
a conjugate quark. Meanwhile, several other studies have confirmed this work
\cite{confirm}.

The above discussion was for $N_c=3$. For $N_c=2$ the situation is
different \cite{different}. Then the Dirac operator is 
selfconjugate leading to a real fermion determinant. 
Because in this case also baryons can be Goldstone bosons,
we expect a critical chemical potential of $\mu_c \sim \sqrt m$.
Mathematically, the Dirac operator for $N_c =2$ has an additional 
anti-unitary symmetry. As is well-known in random matrix theory, the 
anti-unitary symmetries lead to three different universality classes
\cite{dyson,Vprl} characterized by the so called Dyson index $\beta$
($\beta =1$ for real, $\beta =2$ for complex, and $\beta =4$ 
for quaternion real matrix elements). 
In this letter we study a random matrix model of the QCD partition function
at nonzero chemical potential for all three values of $\beta$.

The continuum Euclidean QCD partition function for $N_f$  flavors
with masses $m_f$, and chemical potential $\mu$ can be written as
an average over the Yang-Mills action,
\be
Z(m, \mu) = \left < \prod_{f=1}^{N_f}
\det(\gamma\cdot D +m_f+\mu\gamma_0) \right >_{S_{QCD}},
\label{QCDpart}
\ee
where $\gamma\cdot D$ is the Euclidean Dirac operator and $\gamma_\mu$ 
are Euclidean Dirac matrices.  
In lattice QCD, the chemical potential is incorporated by 
including a factor $e^\mu$ in the forward time links
and a factor $e^{-\mu}$ in the backward  time links. This implementation
does not affect the symmetry relations  discussed below.

To obtain a random matrix theory corresponding to (\ref{QCDpart}) 
we first write the Dirac operator in a chiral basis. Then the
$\mu$ independent nonzero matrix elements are replaced
by Gaussian distributed random variables corresponding to 
the anti-unitary symmetries of the QCD
partition function. In analogy with the random matrix partition function
at zero \cite{SV} and nonzero temperature \cite{JV},
for $N_f$ flavors, this partition 
function is defined as (see \cite{verb} for a review)
\be
Z(m,\mu) = \int D C P(C) \prod_f^{N_f}
\det (D(\mu)+ m_f).
\label{ranpart}
\ee
Here, $C$ is an arbitrary $N\times N$ matrix 
with real, complex or quaternion real 
matrix elements. The integration measure 
$DC$ is the Haar measure.
The probability distribution $P(C)$ is given by
\be
P(C) = \exp (-N (\beta/{2}) {\Sigma^2} {\rm Tr} C C^\dagger) \ .
\label{probal}
\ee
From now on we will work in units where $\Sigma = 1$.

In the case of $N_c \ge 3$ ($\beta= 2$) 
there are no anti-unitary symmetries and the Dirac operator is given by
\cite{Stephanov}
\be
D(\mu) = \left ( \begin{array}{cc} 0 & iC + \mu \\
iC^\dagger +\mu & 0 \end{array} \right). 
\label{chguemu}
\ee
Here, and in (6,8)
below, $\mu$ is proportional to the identity.

For $N_c =2$, the Dirac operator in (\ref{QCDpart}) 
is subject to an additional anti-unitary symmetry \cite{Vprl}. 
We stress that
a nonzero chemical potential does not violate this symmetry.
For continuum fermions (and naive lattice fermions \cite{note})
in the fundamental representation we have
\be
[C\tau_2 K, i\gamma D(\mu)] = 0,
\ee
where $C$ is the charge conjugation matrix, and $i\gamma D(\mu)$ is the 
continuum Dirac operator at $\mu \ne 0$. 
Because $(C\tau_2 K)^2 =1$, it
is possible to find a basis in which the Dirac operator is real.
If we notice that $ \mu\gamma_0$ is Hermitean
we arrive at the following random matrix Dirac operator
\be
D(\mu) = \left ( \begin{array}{cc} 0 & C + \mu\\
-C^T +\mu & 0 \end{array} \right) \ ,
\label{chgoemu}
\ee
where $C$ is an arbitrary real matrix. Quenched instanton simulations
at $\mu \ne  0$ are in this class \cite{Thomas}.

For staggered fermions the gamma matrices are absent and for $N_c=2$ the 
anti-unitary symmetry is given by \cite{Teper},
\be
[\tau_2 K, D^{S}(\mu)] = 0,
\ee
where $D^{S}(\mu)$ is the staggered Dirac operator at $\mu \ne 0$.
Because
$(\tau_2 K)^2 = -1$, we can organize the matrix elements of
the Dirac operator into real quaternions. For an 
(anti-)Hermitean Dirac operator
the anti-unitary symmetry results in a pairwise degeneracy of the eigenvalues. 
This degeneracy
is broken at $\mu \ne 0$. Then the eigenvalues of $D^{S}$ occur in
complex conjugate pairs.
In this  symmetry class with $\beta =4$ the random matrix model is given
by
\be
D(\mu) = \left ( \begin{array}{cc} 0 & C + \mu{\bf 1}\\
-C^\dagger+\mu{\bf 1} & 0 \end{array} \right) \ ,
\label{chgsemu}
\ee
where the matrix elements of $C$ are real quaternions and ${\bf 1}$ is the
unit quaternion. In addition to $D^{S}$ for $N_c = 2$, this class
also contains the Dirac operator for gauge fields in the adjoint
representation \cite{Vprl} for $N_c \ge 2$.

In all three cases the spectral density at $\mu = 0$ is a semicircle which in
the normalization defined by (\ref{probal}) (with $\Sigma= 1$) is given by
\be
\rho_{\rm SC}(\lambda) = \frac 1{2\pi} \sqrt{4-\lambda^2}.
\label{semicircle1}
\ee

The above random matrix models apply to universality classes
in which the eigenfunctions at $\mu = 0$
are $extended$. The eigenvalues corresponding to $localized$ wave functions
are statistically independent with spectral correlations given by 
the Poisson ensemble. A random matrix model in this class without a
chiral structure was recently considered in \cite{Hatano,Efetov}. 
At $\mu \ne  0$, 
it was found that eigenfunctions
with complex eigenvalues are extended whereas eigenfunctions 
with real eigenvalues remain localized as for $\mu  =0$. 
The surprising result of this 
study was that a localization transition was observed in one dimension.
In the present context the extended states can be interpreted as 
the emergence of
a nonzero baryon number density when the eigenvalues scatter 
in the complex plane at $\mu \ne 0$.
Another remarkable result of this study \cite{Hatano} 
was that in the thermodynamic limit
a finite number of eigenvalues remained
on the real axis for $\mu$ below a critical value. Below we will show 
that the fraction of purely imaginary eigenvalues of the model 
(\ref{chgoemu}) with $\beta = 1$ scales as $\sim 1/\sqrt N$ (Our
convention and the conventions of \cite{Hatano,Efetov} for the
eigenvalues differ by a factor $i$.) 
 
However, spectral correlations of the lattice QCD Dirac operator in four 
dimensions are given by the invariant random matrix ensembles \cite{HV}
even for relatively weak coupling. Moreover, explicit calculations of
the inverse participation ratio in an instanton liquid model shows
delocalized Dirac eigenfunctions \cite{james}.
Therefore, it seems that
the models in which the states are localized at $\mu= 0$ are 
inappropriate for QCD applications.

In the remainder of this paper we study
the spectrum of the random matrix model (\ref{ranpart}). 
The generating function is given by
\be
Z = \langle {\det}^n(z-D) {\det}^n( z^* - D^\dagger)\rangle.
\label{zdet}
\ee
The resolvent is defined as $G(z) = -\del_z\log Z/n$ and the spectral
density in the complex plane is given by $\rho(x,y) = \del_{z^*} G(z)/\pi$. 
The quenched approximation is obtained as 
the limit $n\rightarrow 0$ at the end of the calculation. For integer $n$, 
this partition function can be analyzed in standard fashion 
\cite{HVeff,Stephanov}.
First we write the determinants as Grassmann integrals. Then we perform
the average over the random matrices resulting in a 4-fermion interaction.
If we bosonize this interaction and integrate out the fermions we arrive
at a partition function that is amenable to a saddle point approximation.
At $\mu \ne 0$ this analysis was performed by Stephanov for
Hermitean random matrices. The analysis for $\beta=1$ and $\beta=4$
is somewhat more complicated \cite{HVeff}. 
However, in both cases
we succeeded to solve the saddle-point equations with the remarkable result 
that for the normalization defined by (\ref{probal}), 
the solution coincides with the one obtained for $\beta = 2$.
For all values of $\beta$, the resolvent can 
be evaluated for arbitrary integer values of $n$. 
To leading order in $1/N$ we find the result is independent of $n$.
Because of the absolute value of the determinant in (\ref{zdet}) the
generating function is a smooth function of $n$. This guarantees
the existence of the replica limit.

Numerical results for the random matrix ensembles defined in 
(\ref{chguemu},\ref{chgoemu},\ref{chgsemu}) are shown in Fig. 1. We show
results for $\mu = 0.15$ and $\mu = 0.5$.
The dots represent the eigenvalues in the complex plane. 
The full line is the analytical result \cite{Stephanov}
for the boundary of the eigenvalues given by
the algebraic curve $(y^2 +4\mu^2)(\mu^2-x^2)^2 +x^2 = 4\mu^2(\mu^2-x^2)$. For
$\beta =1$ and $\beta = 4$ we observe that the spectral density 
deviates significantly from the saddle-point result.
For $\beta = 1$ 
we find an accumulation of eigenvalues on the imaginary axis whereas for 
$\beta = 4$ we find a depletion of eigenvalues in this domain. 
This depletion can be understood as follows. For $\mu = 0$ all eigenvalues
are doubly degenerate. This degeneracy is broken at $\mu\ne 0$ which produces
the observed repulsion of the eigenvalues. 

The number of purely imaginary eigenvalues appears to scale as $\sqrt N$, 
which explains that this effect is not visible in a leading order saddle 
point analysis. From a perturbative analysis 
of (\ref{zdet}) one obtains a power series in $1/N$.
Clearly, the $\sqrt N$ dependence
requires a truly nonperturbative analysis
of (\ref{ranpart}). Such a $\sqrt N$  scaling behavior is typical
for the regime of weak non-hermiticity first identified by Fyodorov
et al. \cite{fyodorov}.

A similar cut below a 
cloud of eigenvalues was found in instanton liquid simulations for $N_c =2$
at $\mu \ne 0$ \cite{Thomas} and
in a random matrix model of arbitrary real matrices
\cite{Khor}. The depletion of the eigenvalues
along the imaginary axis was observed earlier in lattice QCD simulations
with staggered fermions \cite{baillie}. 
\begin{figure}[ht]
\twocolumn[
{\psfig{figure=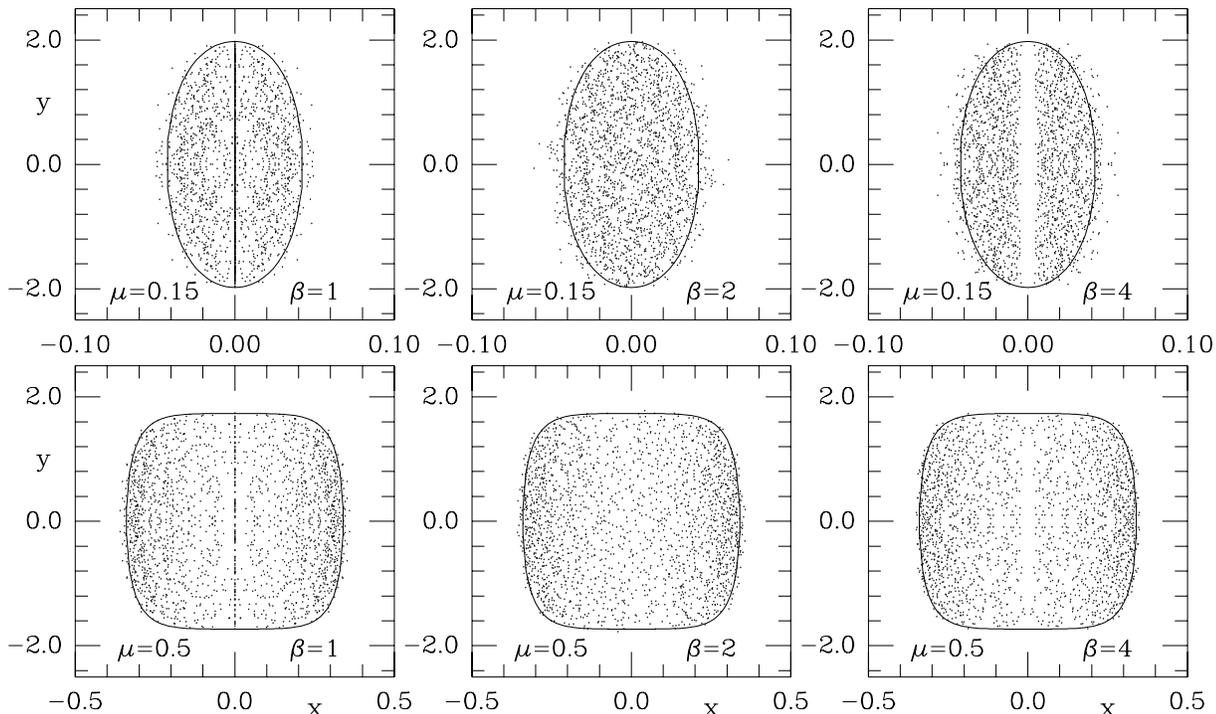,width=160mm,angle=-90}}
\widetext
\caption{
Scatter plot of the real ($x$), and the imaginary 
parts ( $y$) of the eigenvalues of the random matrix Dirac operator.
The values of $\beta$ and $\mu$ are given in the labels of the figure.
The full curve shows the analytical result for the boundary.}
\label{fig1}]
\narrowtext
\end{figure}

In Fig. 2 we plot the fraction of imaginary eigenvalues versus $\mu\sqrt N$ for
$\mu$ equal to 0.05, 0.1 and 0.2 and $N$ ranging from ten  to one thousand. 
The full line represents the analytical result due to Efetov \cite{Efetov}
\be
\alpha_0=
\int_{-2}^2 \rho_{SC}(\lambda) d\lambda \int_0^1 dt 
\exp[-N(2\pi\mu\rho_{SC}(\lambda))^2 t^2],
\ee
where $\rho_{SC}$ is defined in (\ref{semicircle1}). This  result was obtained
for an ensemble
of independent real symmetric matrices perturbed by an anti-symmetric matrix
in the limit that the norm of the perturbing operator is of the order of
the level spacing of the unperturbed matrix (his conventions
differ from ours by a factor $i$). Apparently,  the fraction 
of purely imaginary eigenvalues is not modified by
the chiral structure of the ensemble (\ref{chgoemu}).
Asymptotically, for $\mu \sqrt N \gg 1$, this fraction
is given by $\alpha_0 \sim 1/\mu\sqrt{\pi N}$.

It is well known that the replica trick fails in some cases \cite{martin}. 
For example, it fails for the unquenched partition function 
 \cite{Stephanov}. More typically, it fails in cases where the 
the saddle-point is given by a nontrivial manifold \cite{martin}.
In the limit $\mu \rightarrow 0$ and $ z$ real, the partition
function has a higher degree of symmetry. Therefore, 
in the limit $N\rightarrow \infty$ with $\mu^2 N$ fixed, the solution
of the saddle point equations is given by a nontrivial manifold, and  
there is no guarantee that the replica trick will work .
In order to obtain truly nonperturbative results 
one has to rely on the supersymmetric method for random matrix theory
\cite{efvwz}. This method was extended to non-hermitean 
complex matrices in
\cite{supernonh,fyodorov} and to arbitrary real matrices in \cite{Efetov}.
The application of the supersymmetric
method to the chiral ensembles will eventually provide us with an 
explanation of the scaling behavior of the number of purely imaginary 
eigenvalues (\cite{All}).
\begin{figure}[h]
\centerline{\psfig{figure=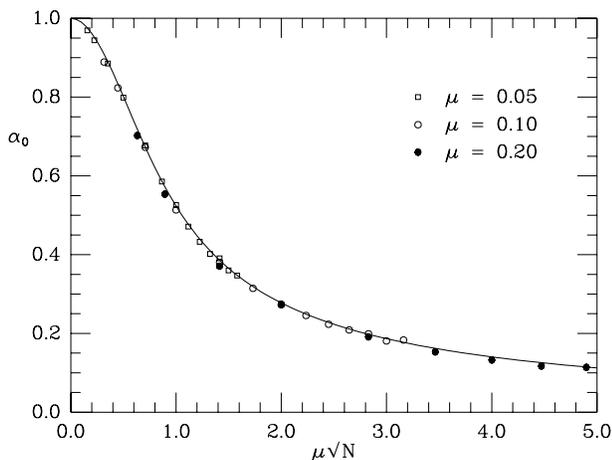,angle=-90,width=80mm}}
\caption{ The fraction of real eigenvalues, $\alpha_0$ 
versus $\mu \sqrt N$ for three different values of 
$\mu$ and matrices ranging from $N= 10$ to $N=1000$. 
The full line shows Efetov's result [12].}
\label{fig2}
\end{figure}

Stephanov has shown analytically that the 
quenched approximation does not work for $\beta =2$. Specifically, the 
unquenched partition function (\ref{ranpart}) for $N_f \ne  0$ 
results in a nonzero
chiral condensate below a critical value of $\mu$. However in the 
quenched case, given by the $n\rightarrow 0$ limit of (\ref{zdet}),
the chiral condensate is zero at any $\mu\ne 0$.
The situation for $\beta = 1$ and $\beta = 4$ is different. Then in both
cases the fermion determinant is real for real $z$. 
(For $\beta = 4$ this follows by using the identity 
$q^* = \sigma_2 q \sigma_2$ for a quaternion real element $q$.)
Therefore, for real $z$, the generating function (\ref{zdet}) 
for $n$ replicas is identical to the partition function (\ref{ranpart})
for $2n$ flavors with mass $z$. As mentioned above, the saddle point
result for the resolvent defined by the generating function (\ref{zdet})
does not depend on $n$ and gives the quenched result for $n\rightarrow 0$.
We thus conclude that quenching works for an even number of flavors.
Consequently, chiral symmetry will be  restored for arbitrarily 
small nonzero $\mu$, whereas a condensate of
a quark and a conjugate quark develops. Indeed, 
this phenomenon has been observed
in the strong coupling limit of lattice QCD 
with two colors \cite{Teper,Elbio}.

In conclusion, at $\mu \ne 0$
we have found that the depletion of Dirac eigenvalues on the
imaginary axis observed in 
lattice QCD simulations and the accumulation of
Dirac eigenvalues found in instanton liquid simulations 
(both for two colors and quenched)
is a generic feature related to the anti-unitary symmetries of the Dirac 
operator. 
If it turns out that 
the Dirac eigenfunctions are localized at $\mu= 0$ 
(see \cite{Janssen} for lattice results in this direction) 
the fraction of purely imaginary eigenvalues at $\mu\ne 0$ might
remain finite in the thermodynamic limit. For two colors, 
this might lead to  unexpected chiral properties in the continuum limit
at $\mu \ne 0$.
Clearly, this possibility deserves further attention.

This work was partially supported by the US DOE grant
DE-FG-88ER40388. Edward Shuryak and Yan Fyodorov are 
thanked for useful discussions.


\begin{references}
\bibitem{Sommer}H.J. Sommers, Phys. Rev. Lett. {\bf 60}, 1895 (1988).
\bibitem{Khor}B. Khoruzhenko, J. Phys. A: Math. Gen. {\bf 29}, L165 (1996). 
\bibitem{fyodorov} Y. Fyodorov, B. Khoruzhenko and H. Sommers, Phys. 
Lett. {\bf A 226}, 46 (1997); cond-mat/9703152.
\bibitem{Hatano} N. Hatano and D.R. Nelson, Phys. Rev. Lett. {\bf 77}, 
570 (1996). 
\bibitem{Efetov}K.B. Efetov, cond-mat/9702091.
\bibitem{Stephanov}M. Stephanov, Phys.\ Rev.\ Lett.\ {\bf 76}, 4472 (1996).
\bibitem{confirm}R. Janik, M. Nowak, G. Papp and I. Zahed, Phys. Rev. Lett.
{\bf 77}, 4816 (1996);  cond-mat/9612240; 
J. Feinberg and A. Zee, cond-mat/9703087;
M. Halasz, A. Jackson and J. Verbaarschot, 
Phys. Lett. {\bf B395}, 293 (1997); hep-lat/9703006.
\bibitem{everybody}I. Barbour {\it et al.}, Nucl.\ Phys.\ {\bf B275}, 296 
(1986); M.P. Lombardo, J. Kogut, and D. Sinclair, hep-lat/9511026.
\bibitem{mendel}W. Wilcox, S. Trendafilov and E. Mendel, Nucl. Phys.
Proc. Supple {\bf 42} (1995) 557.
\bibitem{Gocksch} A. Gocksch, Phys. Rev. Lett. {\bf 61} (1988) 2054.
\bibitem{different}J. Kogut, {\it el al.}, 
Nucl. Phys. {\bf B225 [FS9]} , 93 (1983).
\bibitem{dyson}F.J. Dyson, J. Math. Phys. {\bf 3}, 140 (1962).
\bibitem{Vprl}J. Verbaarschot, Phys. Rev. Lett. {\bf 72}, 2531 (1994).
\bibitem{JV}A.D. Jackson and J. Verbaarschot, Phys.\ Rev.\ {\bf D53}, 7223 
(1996); T. Wettig, A. Sch\"afer and H. Weidenm\"uller,
Phys. Lett. {\bf B367}, 28 (1996).
\bibitem{verb}J. Verbaarschot, Nucl. Phys. Proc. Suppl. {\bf 53}, 88 (1997).
\bibitem{note} For Wilson fermions the anti-unitary symmetry is
$[\gamma_5 C \tau_2 K, \gamma_5 \gamma D^W] = 0$, but the $r$-term violates
the chiral symmetry, $\{\gamma_5, \gamma_5 \gamma D^W\}\ne 0$ (see \cite{HV}). 
\bibitem{SV}E. Shuryak and J. Verbaarschot, Nucl. Phys. {\bf A560},
306 (1993).
\bibitem{Thomas} E. Shuryak and Th. Sch\"afer, {\it private communication}.
\bibitem{Teper}S. Hands and M. Teper, Nucl. Phys. {\bf B347}, 819 (1990).
\bibitem{HV}M. Halasz and J. Verbaarschot, Phys. Rev. Lett.
{\bf 74}, 3920 (1995); M. Halasz, T. Kalkreuter, and J. Verbaarschot,
Nucl. Phys. Proc. Suppl. {\bf 53}, 266 (1997).
\bibitem{james} J.C. Osborn and J. Verbaarschot, in preparation.
\bibitem{HVeff}M.A. Halasz and J. Verbaarschot, Phys. Rev. {\bf D52}, 
2563 (1995).
\bibitem{baillie} C. Baillie, K. Bowler, P. Gibbs, I. Barbour and M. 
Rafique, Phys. Lett. {\bf 197B}, 195 (1987).
\bibitem{martin}J. Verbaarschot and M. Zirnbauer, J. Phys.
{\bf A17}, 1093 (1985).
\bibitem{efvwz}K.Efetov, Adv. Phys. {\bf 32}, 53 (1983);
J. Verbaarschot, H. Weidenm{\"u}ller, and M. Zirnbauer, Phys. Rep.
{\bf 129}, 367 (1985).
\bibitem{supernonh}Y. Fyodorov and H. Sommers, JETP Lett. {\bf 63} (1996), 1026.
\bibitem{All}M.A. Halasz, $et$ $al.$, {\it in preparation}.
\bibitem{Elbio} E. Dagotto, F. Karsch and A. Moreo, Phys. Lett. 
{\bf 169 B}, 421 (1986). 
\bibitem{Janssen}K. Jansen, C. Liu, H. Simma and D. Smith, Nucl. Phys. Proc.
Supp. {\bf 53}, 262 (1997).
\end{references}
\end{document}